# Job Selection in a Network of Autonomous UAVs for Delivery of Goods

Pasquale Grippa[1], Doris A. Behrens[2,3],
Christian Bettstetter[1,4], and Friederike Wall[5]

[1]Institute of Networked and Embedded Systems Alpen-Adria-Universität Klagenfurt, Klagenfurt, Austria

[2]School of Mathematics, Cardiff University 23 Senghennydd Road, Cardiff CF24 4AG

[3]Aneurin Bevan University Health Board Lodge Road, Caerleon NP18 3XQ

[4]Lakeside Labs GmbH, Klagenfurt, Austria

[5]Institute for Business Management, Alpen-Adria-Universität Klagenfurt, Klagenfurt, Austria

Emails: pasquale.grippa@aau.at, BehrensD1@cardiff.ac.uk

## Abstract

This article analyzes two classes of job selection policies that control how a network of autonomous aerial vehicles delivers goods from depots to customers. Customer requests (jobs) occur according to a spatio-temporal stochastic process not known by the system. If job selection uses a policy in which the first job (FJ) is served first, the system may collapse to instability by removing just one vehicle. Policies that serve the nearest job (NJ) first show such threshold behavior only in some settings and can be implemented in a distributed manner. The timing of job selection has significant impact on delivery time and stability for NJ while it has no impact for FJ. Based on these findings we introduce a methodological approach for decision-making support to set up and operate such a system, taking into account the trade-off between monetary cost and service quality. In particular, we compute a lower bound for the infrastructure expenditure required to achieve a certain expected delivery time. The approach includes three time horizons: long-term decisions on the number of depots to deploy in the service area, mid-term decisions on the number of vehicles to use, and short-term decisions on the policy to operate the vehicles.

## 1 Introduction

Small unmanned aerial vehicles (UAVs) have successfully found their way to civil applications. A broad variety of UAV models has been commercialized in the past few years. They fly routes in an autonomous manner, carry cameras for aerial photography, and may transport goods from one place to another. The range of applications is broad, including aerial monitoring of plants and agriculture fields as well as support for first time responders in disasters [21, 35, 27, 1]. Delivering goods via a network of UAVs becomes an option if classical means of transportation — like trucks, trains, and planes — are inappropriate. First, this comes about if roads, railway tracks, or landing facilities do not exist, if weather conditions make it impossible to use them, or if their use is too dangerous or time-consuming. In this context, a compelling service would be the delivery of medicine, vaccinations, or laboratory samples for patients in remote areas and crisis regions. Second, such a service is also worthwhile in densely populated metropolitan areas, when congestion makes roads nearly impassable.

The main objective of this paper is to provide theoretical insight for the architectural setup and control of a UAV-based delivery system. We analyze policies that control how an interconnected team of UAVs resolves service requests that are randomly distributed in space and time.

The entities of the system are goods, customers, vehicles, and depots. Customers request goods that are stored in depots and delivered by vehicles. Service requests, also denoted as jobs or customer demands, are not known in advance and arrive over time at certain locations according to a space-time stochastic process. We analyze both centralized and distributed policies. In the latter, the system's "intelligence" is literally embedded into each vehicle, i.e., each vehicle decides by its own which job to select next, thus raising the autonomy of vehicles to a level that goes beyond autonomous flying.

If the vehicles are capable of consecutively serving several customers before they return to a depot, e.g., if goods are lightweight and total distances are small, the problem of selecting customer requests to be served falls into the domain of dynamic vehicle routing with stochastic demands, dating back to [5, 6, 7]. In contrast to this, vehicles in our system serve no more than one customer per trip for capacity reasons, and we use the term *job selection* to emphasize the difference to routing. By focusing on *non-partitioning* job selection for delivery, we complement research that focused on partitioning routing policies for wide-area surveillance (see [18, 37, 16, 31, 10]).

Our performance measure is delivery time, i.e., the time it takes for a customer from requesting to obtaining a good. The stability of such a service is linked to the queuing of jobs, i.e., customers may have to wait until other customers have received the goods they requested. The system becomes unstable if the average number of waiting customers persistently increases over time. We show features of distributed non-partitioning job selection policies that are found by simulation for $M/G/K$ queues, with $K > 1$, and described in terms of delivery time, which is related to system stability [5, 6, 7]. Using simulations, we deduce some results of general validity on the system behavior. A key insight is that the timing of job selection matters for distributed non-partitioned policies but has no impact on the centralized ones. If, for example, a policy is applied in which the job that first comes in is first served, denoted as FJ-policy, the selection can occur as soon as possible (just after the service of the last cutomer) or as late as possible (just before loading a good). If, in contrast, a policy is applied in which vehicles decide that the currently nearest job is first served, denoted as NJ-policy, the decision on job selection should be made as early as possible. Moreover, we find that a shift in the timing of job selection even alters the structure of the policy in case of NJ-policies (altering also its vulnerability with respect to changes in customer demand).

These insights into system behavior are accompanied by findings about the relation between job selection policies and system setup, the total volume of demand, and system performance. It is known, for example, that FJ-policies outperform NJ-policies for low system loads, while the opposite is true for higher loads, e.g., caused by higher customer arrival rates or a smaller number of vehicles [5]. So far, it is not emphasized in the literature that a system operating according to a FJ-policy will literally collapse all of a sudden and tip into instability if the vehicle fleet is reduced by a single entity. We highlight this threshold effect, as it is essential for both system implementation and operation.

Based on these findings, we derive decision making support for the investment in a delivery system and its operation. In particular, the infrastructure of depots is subject of a long-term decision, the number of vehicles can be modified in mid-term, and the job selection policy is a short-term choice. The set-up of the system (number of depots and vehicles) shapes its monetary costs and, in conjunction with the job selection policy, the service quality provided to the customers for which they are willing to pay for. Hence, for investing in such a delivery system an interesting question is which minimum expenditure is required to provide a certain service quality. For this, we compute a lower bound for the expenditure necessary to set up a stable system as a function of the targeted service quality in terms of average delivery time and exemplarily illustrate the application of this service-possibility-frontier for parameters reported by the company Matternet.

## 2 Related Work

### 2.1 Types of Vehicle Routing Problems (VRPs)

The framework of the used model dates back to [13], who introduced their formulation of the vehicle routing problem (VRP) as a generalization of the traveling salesman problem [17]. Ever since then, the operations research community has intensively studied how a central plan-ner determines optimal sets of routes for fleets of homogeneous vehicles, supplying given sets of geographically dispersed customers with goods [19]. In the context of a "classical VRP", such an optimal set of routes accomplishes that (*i*) all customers are supplied with the demanded products, (*ii*) none of the vehicles exceeds its capacity traveling along its route, (*iii*) no customer is visited more than once, (*iv*) all routes start and end at a central depot, and (*v*) the overall routing cost is minimized. In practical applications, VRPs have a broad diversity of additional requirements and operational constraints affecting the construction of the optimal set of routes. Among these are periodic VRPs [2], VRP with pickup and delivery [14], VRP with split deliveries [15], and VRP with time windows within which customers have to be served [11], to mention but a few. For reviews of exact and approximate methods of solving the classical VRP, we refer to [4, 12, 23, 24, 22, 40, 41], and, for an exhaustive bibliography on vehicle routing to [25].

VRPs are classified according to the nature of system input information. If all system input is known before the vehicles leave the depot(s) and does not change during mission execution, the problem of concern is like the one described in the paragraph above, and said to be both static and deterministic. For many real-world applications, at least some input information, like customer arrivals, behaves according to a probability distribution rather than being known *a priori*. These VRPs are denoted as stochastic. If some input information appears or changes during missing execution, which has to be immediately integrated into decision-making, the VRP is called dynamic ([33, 34], or, for a recent review, [32]). Then, designing sets of routes has to be replaced by designing routing policies, which describe the evolution of motion paths as a function of newly arriving input.

## 2.2 Stochastic and Dynamic Vehicle Routing in Robotics and Aeronautics

The papers [5, 6, 7] were the first to comprehensively analyze the stochastic and dynamic VRP. A generalization of the VRP is the pickup and delivery problem (PDP) [39, 42]. While the VRP was originally investigated for applicability in classical forms of transportation and logistics, there has been a shift towards applications in robotics and aeronautics in the last decade. Particular attention was devoted to the motion coordination of mobile robots, which includes, among others, a VR-based development of spatially-distributed (surveillance) policies for UAVs that are adaptive to network changes [18, 37, 16, 31, 10]. Strategies were developed that ensure that a certain fraction of stochastically generated service requests is served before the jobs expire [29], that account for service priorities [38] and translating demands [8], and that accomplish an effective system management without explicit communication [3]. These approaches utilize partitioning policies to find ways to operate a distributed system that are scalable to large-vehicle networks [31]. These partitioning policies applied to the control of dynamic and stochastic VRPs methodologically differ from distributed partitioning algorithms [30] that analyze how to partition the service area prior to applying a control rule.

We deviate from both of above approaches and seek to find policies that are adaptive in an alternative way. For our application, not partitioning the service area (since depots are fixed) allows for greater system adaptability if vehicles need to endogenously pool in "hot spot" regions of the service area, e.g., whenever disasters or diseases shift demands to certain regions, and if depots run out of goods. Especially in the latter case, non-partitioning job selection policies allow for regarding fixed depots as "customers" that have to be supplied with whatever it is that is needed.

# 3 System Model

## 3.1 Entities of a UAV Delivery System

The system is composed of $K \in \mathbb{N}$ vehicles moving in a bounded and convex service area $\mathcal{A} \subset \mathbb{R}^2$ of size $A := \|\mathcal{A}\|$, where $\| \cdot \|$ is the Euclidean norm. A vehicle is denoted by $v_k$ with identifier $k \in \{1, \dots, K\}$. The current position of vehicle $v_k$ at time $t$ is $\mathbf{v}_k(t) \in \mathcal{A}$ with $t \geq 0$. All vehicles travel at the same constant velocity $\nu \in \mathbb{R}^+$ and are equipped with a battery, whose level at time $t$ is represented by $b_k(t) \in [0, 1]$. The fact that batteries have to be recharged or exchanged is quantified by the parameter $\alpha \in (0, 1]$, which is the air-time ratio, i.e., $\alpha = $ air time/(air time + charge time).

The arrival of customer demands, i.e., delivery requests for goods within $\mathcal{A}$, is generated by a Poisson process with finite intensity $\lambda \in \mathbb{R}^+$, where $\lambda$ is the customer arrival rate. The demands, also called jobs, are indexed by the job identifier $n \in \mathbb{N}$, which indicates the order of request arrivals. The corresponding customer is called $c_n$; his or her position is denoted by $\mathbf{c}_n \in \mathcal{A}$ and assumed to be independently and uniformly distributed in $\mathcal{A}$. Customer demands do not only differ with respect to timing and locations but also with respect to the goods that are requested to be delivered.

Goods are, in general, different but have the same expiration date and are treated with identical priority. There are $L \in \mathbb{N}$ depots in the system, which store the goods. The depots are interconnected and provide a sufficient number of all goods and service activities, like recharging batteries. We assume that, for capacity reasons, a vehicle cannot serve more than one customer demand per trip (see [6, p. 71]). The depots are set up at locations $\mathbf{d} = [\mathbf{d}_1 \ \mathbf{d}_2 \ ... \ \mathbf{d}_L] \in \mathcal{A}^L$, where $\mathbf{d}$ is chosen such that the expected distances between a random point $\mathbf{q} \in \mathcal{A}$ (potential demand) generated according to a uniform distribution over $\mathcal{A}$ and the closest depot are minimal:

$$H_L(\mathbf{d}, \mathcal{A}) := \frac{1}{A} \cdot \int_{\mathcal{A}} \min_{l:l \in \{1,...,L\}} \|\mathbf{d}_l - \mathbf{q}\| d\mathbf{q}. \quad (1)$$

This corresponds to the solution of the continuous multimedian problem known from geometric optimization [28, 44]):

$$\mathbf{d}^* = \arg \min_{\mathbf{d} \in \mathcal{A}^L} H_L(\mathbf{d}, \mathcal{A}), \quad (2)$$

$$\text{with} \quad H_L^*(\mathcal{A}) := H_L(\mathbf{d}^*, \mathcal{A}). \quad (3)$$

A depot is a storage but not a permanent home base for particular vehicles. Whenever a vehicle has delivered a good, it either approaches the nearest depot or another one that is more appropriate to handle the next customer demand. Appropriateness is determined by the job selection rule that is implemented (see Section 5). This non-restricted movement of all $K$ vehicles to all corners of an $L$-depot system is a major difference to related work in [5, 6, 7, 18, 34, 37, 16, 31, 10].

### 3.2 Service Operations and Delivery Time

The delivery time for the customer $c_n$ ($n \in \mathbb{N}$) is denoted by the stochastic variable $T_n = W_n + R_n + S_n$, where $W_n$ is the waiting time, $S_n$ is the service time, and $R_n$ is the return time. Fig. 1 illustrates all operations involved in the service of $c_n$ from arrival at $t = \tau_0$ to the service completion at $t = \tau_{10}$. The waiting time is $W_n = \tau_5 - \tau_0$. The return time is $R_n = \tau_7 - \tau_5$, and depends on the position of the customer $c_{n^-}$ being served before $c_n$ and on the system's load. $R_n$ is included in $[0, R'_n]$: It is null if the vehicle is ready at the depot, and maximum if the service of $c_{n^-}$ is not completed at the arrival of $c_n$. The service time is $S_n = \tau_{10} - \tau_7$. This definition differs Bertsimas and van Ryzin's on-site service time [5, 6, 7] which corresponds to the time a vehicle is "unloading" goods at a customer. It also differs from the notion of service time $B_n$ in classical queuing theory: $S_n = B_n - R_n$.

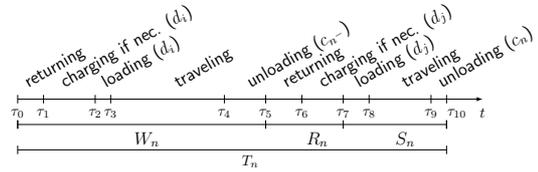

Figure 1: Time intervals involved in a customer service: delivery time $T$, waiting time $W$, return time $R$, and service time $S$.

### 3.3 Queuing Phenomena and Stability

A job selection policy, called $\pi = (\pi_1, \pi_2, ..., \pi_K)$, has to restrain the outstanding jobs [5, 6, 7, 18, 37, 16, 31, 10]. Policy $\pi$ is said to be stabilizing if the expected number of pending jobs (customers waiting for service) stays confined over time, i.e., if there exists an arbitrary constant $\kappa < \infty$ such that

$$\bar{N}_\pi := \lim_{t \to \infty} \mathbb{E}[N(t) | \pi] \leq \kappa, \quad (4)$$

where $N(t)$ denotes the number of pending jobs at time $t$. We assume that $N(0) = 0$, i.e., no customer is waiting at $t = 0$.

The return and service times are crucial for the stability of the system. A necessary condition for stability is [6, p. 63]:

$$\frac{\bar{D}}{\nu} \leq \frac{K}{\lambda} \quad (5)$$

if the on-site service time is null. $\bar{D}$ is the average Euclidean distance between two customers served in sequence, $\lambda$ is the arrival intensity, and $\nu$ is the vehicle speed. This condition applies to our problem with two changes: First, the distance becomes the distance function customer-depot-customer, which divided by the speed gives $\bar{R}' + \bar{S}$ (Fig. 1), where $\bar{R}'$ is the average return time in high load and steady state, and $\bar{S}$ is the average service time in steady state. Second, the effective number of used vehicles is on average $\alpha K$, where $\alpha$ is the airtime ratio. Therefore, our stability condition is

$$\bar{R}' + \bar{S} \leq \frac{\alpha K}{\lambda} \ . \qquad (6)$$

The problem analyzed in this paper can be modeled as an $M/G/K$ queue with interdependent service times $B_n$. $M$ indicates Poisson distributed customer arrivals, $G$ indicates service times distributed according to a generic distribution, and $K$ is the number of servers. For $M/G/K$ queues with independent service times [20], the load factor is defined as

$$\rho \coloneqq \frac{\lambda \bar{B}}{\alpha K} \ . \qquad (7)$$

The system is said to be in light load condition if $\rho \to 0$ and in heavy load condition if $\rho \to 1$. A necessary condition for the stability of the system is $\rho < 1$. In case of stability, $\rho$ can be interpreted as the expected value of the fraction of busy servers. This definition does not apply to our case because $\bar{B}$ depends on the system state [20], which includes the number of waiting customers. Nevertheless, the condition of stability is still valid: $\bar{B}$ approaches $\bar{R}' + \bar{S}$ for $\rho \to 1$, leading to (6). In words, to stabilize the system, it is necessary that the average time between two successfully completed service requests is not larger than the average time between two customer arrivals multiplied by the average number of available vehicles.

## 4 Expenditure for Minimum Infrastructure

We start to derive the minimum expenditure for infrastructure necessary to build a stable system as a function of system performance by deriving a lower bound on average delivery time, $\bar{T}$. This time is bounded by the minimum service time, $\bar{T} \geq \bar{S}_{\min}$. The latter can be expressed in terms of the multi-median function (3), i.e., $\bar{S}_{\min} = H_L^*(\mathcal{A})/\nu$. $H_L^*(\mathcal{A})$ can be bounded by $2\sqrt{A}/3\sqrt{\pi L}$ [44], which implies that

$$\bar{T}_{\min} \geq \frac{H_L^*(\mathcal{A})}{\nu} > \frac{2}{3\nu}\sqrt{\frac{A}{\pi L}}. \qquad (8)$$

By rephrasing (8) we derive a condition for the minimum number of depots necessary to yield a certain average delivery time (larger than minimum time), i.e.,

$$L \geq \frac{4A}{9\pi\nu^2 \bar{T}_{\min}^2}. \qquad (9)$$

From the definition of the load factor (given by (7)) and in consideration of $\bar{S}, \bar{R} \geq \bar{T}_{\min}$ together with $\rho < 1$, we derive a condition for the minimum number of vehicles necessary to yield a certain average delivery time, i.e.,

$$K > \frac{2\lambda}{\alpha} \cdot \bar{T}_{\min}. \qquad (10)$$

For $C_v$ and $C_d$ denoting the costs of a vehicle and a depot, respectively, total infrastructure expenditure is determined by

$$C(K, L) = C_d \cdot L + C_v \cdot K. \qquad (11)$$

With the necessary conditions for a "stabilizing" infrastructure required to yield a particular delivery time, given by (9) and (10), we use (11) to compute an auxiliary function, i.e.,

$$g(\bar{T}_{\min}) = C_d \cdot \frac{4A}{9\pi\nu^2 \bar{T}_{\min}^2} + C_v \cdot \frac{2\lambda}{\alpha} \cdot \bar{T}_{\min}, \qquad (12)$$

and intend to develop a lower bound for total infrastructure expenditure as a function of $\bar{T}$. In this context, it is important to make aware that the values of $\bar{T}_{\min}$ in (12) are contained in a countable set: one value for every $L$, i.e., $\bar{T}_{\min} = \bar{T}_{\min}(L)$. If we changed $\bar{T}_{\min}$ into a continuous variable, $\bar{\tau} \in \mathbb{R}$, the auxiliary function $g$ would have a minimum at

$$\bar{\tau}^* = \sqrt[3]{\frac{4\alpha A \cdot C_d}{9\pi\lambda\nu \cdot C_v}}. \qquad (13)$$

Assume that there exists an $L = L'$ for which the corresponding minimum delivery time is larger than the one defined by (13), i.e., $\bar{T}'_{\min} := \bar{T}_{\min}(L') > \bar{\tau}^*$. Then, the configuration $L = L'$, with $L' < L''$ but $\bar{T}'_{\min} > \bar{T}''_{\min}$, may lead to a higher total expenditure for infrastructure than the configuration $L = L''$. I.e., decreasing the number of depots increases minimum delivery time and may, therefore, increase the number of vehicles necessary to stabilize the system. Depending on the ratio between $C_d$ and $C_v$ this may increase total infrastructure expenditure necessary to build a stable system, denoted by $I_{\min}$. Yet, $I_{\min}$ has to be a non-increasing function of $\bar{T}$. By construction this is accomplished in the following way:

For any two configurations of depots such that $L' < L''$, $\bar{T}'_{\min} > \bar{T}''_{\min}$, and $g(L') > g(L'')$, we are able to obtain delivery time $\bar{T}'_{\min}$ and reduce total expenditure by employing $L = L''$ instead of $L = L'$ depots alongside delaying any delivery by $\bar{T}'_{\min} - \bar{T}''_{\min}$ time units. This makes $I_{\min}$ a piecewise constant function. Additionally, our construction of $I_{\min}$ has to account for $2\lambda \cdot \bar{T}_{\min}/\alpha$ being an integer. Altogether, this yields

$$I_{\min}(\bar{T}) = \min_{l:l \in \{L,\dots,\infty\}} C_d \cdot \frac{4A}{9\pi(H_l^*(\mathcal{A}))^2} + C_v \cdot \left\lceil \frac{2\lambda H_l^*(\mathcal{A})}{\alpha \nu} \right\rceil$$

$$\text{for } \begin{cases} \frac{H_L^*(\mathcal{A})}{\nu} \leq \bar{T} < \frac{H_{L-1}^*(\mathcal{A})}{\nu} & L = 2, 3, \cdots \\ \frac{H_L^*(\mathcal{A})}{\nu} \leq \bar{T} & L = 1 \end{cases} \quad (14)$$

which constitutes the minimum infrastructure expenditure *necessary* for building a stable system that meets a targeted performance $\bar{T}$. Note that (14) does not constitute the amount of financial resources *sufficient* to guarantee system stability because the bound on delivery time is not tight for $\rho \to 1$, and, in general, policies do not stabilize the system for all $\rho < 1$.

# 5 Job Selection Policies

Job selection goes beyond merely picking the next demand. It has to specify all decisions needed to operate a delivery system, including which customer demand to serve first, which vehicle to let serve the next customer demand, at which depot to let vehicles load up goods, which paths to let vehicles follow, and where to let vehicles return to if no customers are waiting. We analyze two classes of job selection policies: nearest job first (NJ) and first job first (FJ). NJ-policies select jobs based on the location of the customer; FJ-policies select jobs based on the arrival time of the customer's job. By comparing these two policies, we seek to gain insight to the following question: Is it worth to delay the decision on job selection to obtain more information about new customer requests in order to reduce the delivery time? For both classes we evaluate two extreme cases: when selection is made as soon as possible (just after the previous service), and when selection is made as late as possible (just before loading the good).

A first issue in this context is whether the vehicles consider the arrival order of customers in the waiting queue. A second issue is the *timing of job selection*. In particular, it seems plausible to assume that decisions should be made "as late as possible" to utilize the most recent information about the system status, basically postponing the individual job selection until loading goods in the depot. Such delayed decisions could also have a negative effect. This relates to the fact that, for $L \geq 2$, one of the most important decisions to be made is where a vehicle should return to after satisfying a customer demand. If a non-postponed decision about the next job includes a "clever" selection of the depot to travel to, while a postponed decision leads to being at a suboptimal place at the moment of job selection, the advantage gained by processing more recent information is counteracted by the disadvantage of the extra distance traveled to the customer to be served next. We will investigate this effect in Section 7.

A third issue is the *coordination of job selections* to avoid that more than one vehicle selects the same job. Such a scenario where a particular customer, say $c_{\tilde{n}}$, is selected by $K'$ vehicles with $2 \leq K' \leq K$, can easily happen if the system is not fully utilized, i.e., for $\rho \to 0$. FJ-policies are centralized, i.e. there is an inherent coordination mechanism among vehicles. We refer to this mechanism as assortative: a central entity selects the next job and assigns it to the nearest vehicle. Hence, the selection is based on time arrival and positions of waiting customers, positions of depots, and positions of all vehicles. NJ-policies are distributed. Every vehicle makes selections based on its own position with respect to the positions of customers and depots, and whenever a vehicle selects a job this is removed form the list of waiting jobs. Every vehicle only knows its own position, the positions of depots, and the positions of waiting customers. For these policies, the coordination mechanism is randomized: Every time that more than one vehicle has the same time of job selection, a random priority order is assigned to the vehicles involved. The randomize mechanism corresponds to a real case where vehicles are not coordinated at all and the connection delay to the list of waiting customers is not predictable because it depends on the network conditions.

Table 1 shows the job selection policies used in this article and classifies them according to three features: the order of jobs to be done, the timing of job selection, and the coordination of jobs in case of ambiguous selection. 1) Policies in which vehicles select jobs according to their current distance from the customer can be named "nearest job first served" and are called NJ-policies. Policies in which vehicles select the smallest $n$ with $c_n \in \mathcal{N}(t)$ (where $t$ indicates the instance of job selection) are named "first job first served" and are called FJ-policies. 2) If vehicles make their individual job selection immediately after completion of a service, the policy is marked with subscript $+$. If vehicles make their job selection later in time, when they are ready to upload goods at a depot, the policy is marked with a subscript $-$. 3) The subscripts $r$ and $a$ correspond to random and assortative job coordination in case of an ambiguous job selection process, respectively. The following four policies are used as reasonable combinations: $\pi^{\mathrm{NJ}}_{+r}$, $\pi^{\mathrm{NJ}}_{-r}$, $\pi^{\mathrm{FJ}}_{+a}$, and $\pi^{\mathrm{FJ}}_{-a}$. A feature included in all policies is the effect of a vehicle $v_k$'s battery level, $b_k(t)$, on whether or not a vehicle selects a job at time $t$ at all. In particular, if at the instance of job selection the battery level is less than $30\,\%$ the vehicle approaches the nearest depot to recharge and does not select a job until its battery level has again reached $80\,\%$. If there is no newly upcoming service request, $v_k$ goes on with charging until the battery is full.

## 5.1 NJ-Policies With Randomized Job Coordination

Policy $\pi^{\mathrm{NJ}}_{+r}$ is called *Do Nearest Job*. After completion of a service, the vehicle $v_k$ selects the next customer such that the travel distance to this customer via a generic depot is minimized. In mathematical terms, we solve

$$\min_{\substack{l:l\in\{1,...,L\}\\ n:c_n\in\mathcal{N}(\tau)}} \|\mathbf{v}_k(\tau)-\mathbf{d}_l\| + \|\mathbf{d}_l-\mathbf{c}_n\|, \quad (15)$$

where $\tau$ is the time instant of completion of the previous job, which is here equal to the time instant of the new job selection. Such selection is made if the battery level at time $\tau$ is large enough, i.e., $b_k(\tau) \geq 0.3$; otherwise a vehicle will return to the nearest depot. Every vehicle has to make $L \cdot N(\tau)$ comparisons to conclude individual job selection. If a job minimizes the traveling distance for more than one vehicle, according to the random coordination described above, only one vehicle randomly chosen will serve the job.

Policy $\pi^{\mathrm{NJ}}_{-r}$ is called *Rush to Depots*. After completion of a service, the vehicle $v_k$ first travels to the closest depot. The vehicle then selects the nearest customer, i.e., it solves:

$$\min_{n:c_n\in\mathcal{N}(\tau+R)} \|\mathbf{v}_k(\tau+R)-\mathbf{c}_n\|, \quad (16)$$

where $\tau+R$ is the time instant $v_k$ arrives at the depot. The vehicle remains in the depot if there is no waiting customer. If a job minimizes the traveling distance for more than one vehicle, the supplying vehicle is determined randomly. This policy requires $L+N(\tau+R)$ comparisons.

## 5.2 FJ-Policies With Assortative Job Coordination

Following a first come first served (FCFS) policy, a central unit always selects the "oldest" demand request still unaddressed, i.e., the smallest $n$ with $c_n \in \mathcal{N}(t)$, where $t$ indicates the instance of job selection. Let this demand be called $\underline{n}(t)$. In particular, policy $\pi^{\mathrm{FJ}}_{+a}$ (also called *FCFS by Nearest Vehicle*; see Tab. 1) implies that whenever one or more than one vehicles complete a service, at time $t=\tau$, a central controller selects the next customer in line, i.e., demand $\underline{n}(\tau)$, and chooses the vehicle $v_{k^*}$ and the path through the depot, $d_{l^*}$, that minimizes the total distance between its current position $\mathbf{v}_k(\tau)$, a depot's position $\mathbf{d}_l$, and the customer's position $\mathbf{c}_{\underline{n}(\tau)}$:

$$\min_{\substack{l:l\in\{1,...,L\}\\ k:k\in\mathcal{K}'}} \|\mathbf{v}_k(\tau)-\mathbf{d}_l\| + \|\mathbf{d}_l-\mathbf{c}_{\underline{n}(\tau)}\|, \quad (17)$$

where $\mathcal{K}'$ is the number of vehicles ready for the service. Obeying (17) a total of $K' \cdot L$ computations is needed to come up with a vehicle's individual selection of whom to next serve.

Obeying the rules of policy $\pi^{\mathrm{FJ}}_{-a}$ (or *FCFS by First Vehicle at Depot*; see Tab. 1), whenever one or more than one vehicle is available to a depot, the central unit selects the demand $\underline{n}(\tau + R)$ and performs $K'$ computations to determine the vehicle nearest to that job. Subsequently $L$ computations are performed to find the depot nearest to the selected customer where the vehicle has to return to after the delivery:

$$\min_{k:k\in\mathcal{K}'} \|\mathbf{v}_k(\tau + R) - \mathbf{c}_{\underline{n}(\tau + R)}\| + \min_{l:l\in\{1,\ldots,L\}} \|\mathbf{c}_{\underline{n}(\tau + R)} - \mathbf{d}_l\|. \quad (18)$$

**Table 1: Features of Job Selection Policies**

| Features | Do Nearest Job ($\pi_{+r}^{\text{NJ}}$) | Rush to Depots ($\pi_{-r}^{\text{NJ}}$) | FCFS by Nearest Vehicle ($\pi_{+a}^{\text{FJ}}$) | FCFS by First Vehicle at Depot ($\pi_{-a}^{\text{FJ}}$) |
|---|---|---|---|---|
| Jobs are selected in first come first served (FCFS) order from a shared queue. | - | - | Yes | Yes |
| There is no specific selection order of jobs. | Yes | Yes | - | - |
| Decisions are made at a customer, just after completing a service. | Yes | - | Yes | - |
| Decisions are made at a depot, just before loading goods. | - | Yes | - | Yes |
| If more than one vehicle seeks to select a job, a random one does. | Yes | Yes | - | - |
| If more than one vehicle seeks to select a job, the nearest one does. | - | - | Yes | Yes |
| If there are no jobs a vehicle goes to the nearest depot. | Yes | Yes | Yes | Yes |
| If a vehicle's battery level is less that 30% the vehicle approaches the nearest depot. If due to recharging a vehicle's battery reaches a level higher than 80% the vehicle is ready again for service. If there is no customer the vehicle fully recharges its battery. | Yes | Yes | Yes | Yes |

# 6 Simulation Setup

For $\pi \in \{\pi_{+r}^{\text{NJ}}, \pi_{-r}^{\text{NJ}}, \pi_{+a}^{\text{FJ}}, \pi_{-a}^{\text{FJ}}\}$, we simulate the movement of $K \in \{1, \ldots, 24\}$ vehicles in a square area of $A = 16$ km$^2$. The $L \in \{1, 4, 9, 16\}$ depots are located such that they minimize the average distance of any potential demand in $\mathcal{A}$ from the nearest depot. Customer demands are assumed to randomly arrive over time at a rate $\lambda = 0.65$ requests/minute. We consider the non-variation of $\lambda$ as feasible, since with $K$ we vary $\lambda$'s counterpart in affecting the load factor, defined by (7), so to derive sufficient information about the performance of the system for varying levels of resource utilization.

All vehicles travel at a constant velocity of $\nu = 30$ km/h, neglecting acceleration and deceleration phases and neglecting the extra time for starting and landing. Vehicles' air time are limited due to their finite battery capacity, which is assumed to be full at $t = 0$. We choose the ratio between air time and charge time as being $1/3$, i.e., $\alpha = 0.25$. Moreover, we assume that at $t = 0$ every $v_k$ is placed at one of the depots. Finally, we neglect the loading time (goods).

In the simulations, the system presented in Section 3 is observed every $\delta$ time units. Interarrival times are generated according to a geometric distribution so that the number of customer arrivals per time unit follows a binomial distribution with parameters $h = 1/\delta$ and $p = \lambda\delta$. The Poisson distribution is a sufficient approximation of the binomial distribution if $p \leq 0.08$ and $h \geq 1500p$ [9]. Therefore, for all of our simulation runs, we choose $\delta$ such that $\delta \leq \min\{0.08/\lambda, 1/\sqrt{1500\lambda}\}$.

The data collected from the simulations are used to compute the average delivery time. Specifically, we firstly estimate the length of the warm-up phase using Welch's method [43]. Secondly, we compute the average delivery time with the replication/deletion method [26], which is equivalent to the independent replications method [43]. These methods are used for statistical analysis of data and enable us to estimate expected values and confidence intervals.

For every parameter setup, we perform 10 simulations with a minimum of 4,000 customer requests for every simulation. If it is impossible to evaluate whether the system reaches the steady state within this period, we simulate 10,000 customers requests, which is more than 10 days of continuous operation.

# 7 Simulation Results

## 7.1 Warm-Up Phase

We discuss the computation of the warm-up phase for the policies *Do Nearest Job* and *FCFS by Nearest Vehicle* for $K \in \{11, 12, 14, 16\}$ vehicles and $L = 16$ depots.

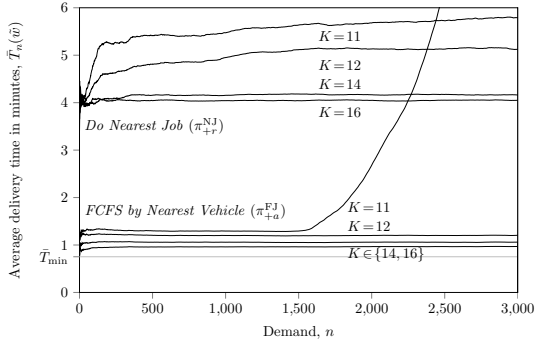

Figure 2: Delivery time averaged over 10 simulation runs and $2\tilde{w} + 1 = 1001$ demands vs. demand index $n$ for $K$ vehicles.

Fig. 2 reveals that the length of the transient phase decreases with an increasing number of vehicles, i.e., with a decreasing load factor. We conclude that delivery systems provided with fewer vehicles must be simulated for a longer period (higher number of demands $\tilde{n}$) to reach steady state conditions. Being conservative, we estimate that the length of the warm-up phase is $\tilde{n}_{\mathrm{wu}} = 3000$ for the case $K = 11$, while being $\tilde{n}_{\mathrm{wu}} = 2000$ for the case $K = 12$, and $\tilde{n}_{\mathrm{wu}} = 500$ for all other cases.

Moreover, Fig. 2 introduces a result addressed when discussing Fig. 3d below: *FCFS by Nearest Vehicle* ($\pi_{+a}^{\mathrm{FJ}}$) can yield a substantially better system performance than *Do Nearest Job* ($\pi_{+r}^{\mathrm{NJ}}$). The only drawback observed is that, while $\pi_{+a}^{\mathrm{FJ}}$ gets very close to the minimum average delivery time $\bar{T}_{\min}$ for $K \in \{12, 14, 16\}$, it cannot stabilize the system for $K = 11$.

## 7.2 Evaluating Job Selection Policies

Fig. 3 shows the average delivery time $\bar{T}$ for different job selection policies as a function of the number of vehicles $K$ and the number of depots $L$. The shaded areas indicate "impossible regions", i.e., all ordered pairs $(K, \bar{T})$ for which either $\rho > 1$ or $\bar{T} < \bar{T}_{\min}$ or both. Depot numbers are in the set $\{2^n; \ n = 0, 1, 2, 4\}$ so that the multi-median problem has traceable mathematical solution for a squared service area: The area is divided into $L$ identical squares, and the depots are located in the centroids of the squares. The following basic phenomena can be observed.

**Tipping Point Behavior** Increasing the number of vehicles $K$ has no effect on system performance once $K$ exceeds a certain threshold $\tilde{K}$. This observation holds for all studied policies. It results from the fact that, for any $K$ larger than $\tilde{K}$, there is a vehicle in the chosen depot ready to serve a new demand, such that the overall delivery time $T$ just consists of the service time $S$. In such light load conditions, FJ-policies ($\pi_{+a}^{\mathrm{FJ}}$ and $\pi_{-a}^{\mathrm{FJ}}$) yield better system performance than NJ-policies ($\pi_{+r}^{\mathrm{NJ}}$ and $\pi_{-r}^{\mathrm{NJ}}$). Such policy differences in system performance are known from [5, 6], and they are amplified by the differences in job coordination in case of ambiguity.

Decreasing the number of vehicles $K$ improves resource utilization but in turn increases the load factor, which eventually implies system instability. The delivery time of *Do Nearest Job* ($\pi_{+r}^{\mathrm{NJ}}$) increases slowly with a decreasing $K$, i.e., the system shows a gradual transition from the region of system underuse to the region of instability. In contrast to this, for all other policies, the system tips into instability all of a sudden. This tipping point behavior of $\pi_{-r}^{\mathrm{NJ}}$, $\pi_{+a}^{\mathrm{FJ}}$, and $\pi_{-a}^{\mathrm{FJ}}$ can be explained by the fact that these policies resemble the behavior of the $K$ stochastic queue median ($K$-SQM) policy introduced by [6, p. 65], while $\pi_{+r}^{\mathrm{NJ}}$ in heavy load behaves similar to the nearest neighbor (NN) policy presented by [5, p. 612].

**Timing Matters** The timing of the job selection decision has an impact on delivery time and system stability. Let us compare two extreme cases for both FJ and NJ-policies: the decision is made as soon as possible (just upon the service at the previous customer), or the decision is made as late as possible (just before loading the good at the depot).

Using FJ-policies, timing has no effect on delivery time and system stability (see Fig. 3). In contrast, using NJ-policies, an early decision is beneficial for stability, i.e., *Do Nearest Job* $\pi_{+r}^{\mathrm{NJ}}$ is more robust than *Rush To Depots* $\pi_{-r}^{\mathrm{NJ}}$. The positive effect of more recent information outweighs the negative effect of returning to a depot located sub-optimally (on average) for the next job. The gap between the two NJ-policies widens with more depots, i.e., the effect of "being at a suboptimal place" is more pronounced in systems with many depots. This happens be-cause, in heavy load conditions, randomized job coordination is rarely necessary (if at all), and $\pi_{+r}^{\text{NJ}}$ behaves similar to a NN policy (as mentioned above), which serves the nearest demand available after every service completion.

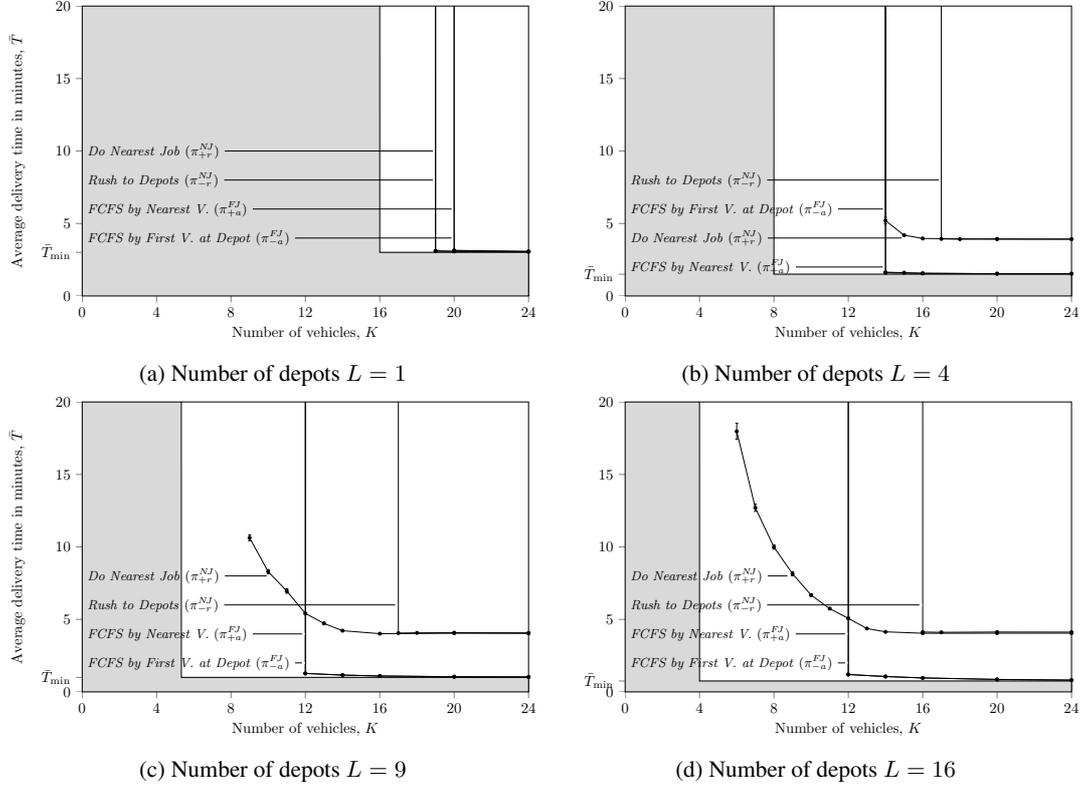

(a) Number of depots $L = 1$

(b) Number of depots $L = 4$

(c) Number of depots $L = 9$

(d) Number of depots $L = 16$

Figure 3: Average delivery time (with 90% confidence) vs. number of vehicles for $L$ depots.

In summary, the timing of job selection has fundamental impact on the qualitative behavior of NJ-policies: it changes the vulnerability with respect to variation in customer demand. This phenomenon does not occur for our FJ-policies: both maintain structurally equivalent to K-SQM policies.

## 7.3 Decision Making Support

We have seen that the number of depots must be chosen in relation to the size of the service area, and that this long-term choice on depot infrastructure must be coordinated with the mid-term choice on vehicles and the short-term choice on the job selection policy. For company-specific parameter values, a diagram like Fig. 4 translates the insights derived in the subsection above into the monetary domain. It relates a company's expenditure $I$ for depots and vehicles to average delivery time $\bar{T}$. The purpose of such a plot is to provide decision making support for companies that set up an airborne delivery system equipped with small UAVs.

To give an example, Fig. 4 is produced on the assumption that the cost of a UAV suitable to deliver two-kilogram packages is 1,000 US\$ plus a maintenance cost of 100 US\$ per annum, and the cost of a depot is 15,000 US\$ plus a maintenance cost of 500 US\$ per annum. Operating the system over ten years, the costs per vehicle and depots are $C_v = 2,000$ US\$ and $C_d = 20,000$ US\$, respectively. These parameter values are those assumed by the company Matternet [36].

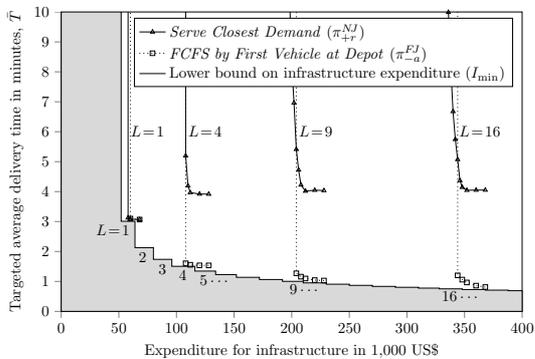

Figure 4: Relation between infrastructure expenditure and average delivery time; $C_v = 2,000$ US\$, $C_d = 20,000$ US\$.

A lower bound on the expenditure required to build a stable system, denoted by $I_{\min}$, is derived in Sec. 4 and is given by (14). Fig. 4 plots this bound for the given parameters, where all values below the bound are shown as a shaded area. The bound has a "staircase" shape with tread levels being the minimum average delivery time achievable with a particular number of depots. No operable system exists for parameters in this area, while every combination of $I$ and targeted $\bar{T}$ located above fulfills $\rho < 1$ and $\bar{T} > \bar{T}_{\min}$. The bound corresponds to a *service possibility frontier*; it gives a necessary but not a sufficient condition for infrastructure expenditure. The actual performance is, policy depended, and more financial resources than $I_{\min}$ may be needed to operate the system in a stable manner and to meet the targeted performance.

A company that wants to operate a delivery service and serve a customer within a certain average delivery time can employ such a diagram as follows: If the average delivery time should be no more than $\tau$, the company has to look for feasible combinations of infrastructure and stabilizing policies, i.e., squares and triangles, that are located as close as possible to the origin and below the $(\bar{T} = \tau)$-line. If the customers' willingness to pay for several levels of service quality is given, it is possible to quantify the company's marginal revenues of increasing performance. From Fig. 4 we know the marginal cost of decreasing delivery time. Then, we are able to determine the combination of infrastructure and job selection policy that maximizes the company's profit. For the parameters of Matternet we find that a system with $L = 1$ depot can serve a customer in about three minutes on average in an area of 16 km$^2$, which comes at a cost of 60,000 US\$. This time can be reduced to less than 1.5 minutes if the company spends more than 100,000 US\$ for infrastructure (associated with $L = 4$ depots). If the company's marginal revenue of reducing delivery time by 1.5 minutes (50 %) is larger than approximately 40,000 US\$, the four-depot configuration is better than the one-depot configuration. In other words, Fig. 4 informs about the financial resources required for achieving a certain quality of service with a certain policy and about the volume of additional monetary resources obligatory to "buy" a shorter delivery time.

# 8 Conclusions

We addressed the architectural setup and mechanisms of self-control in a network of UAVs for delivery of goods. We combine short-term decisions (e.g., job selection policy), mid-term decisions (e.g., number of vehicles), and long-term decisions (e.g., number of depots) into an integrated decision making model. Taking into account the short and mid-term horizons, the delivery time of the system experiences a threshold behavior: a stable system could lead to instability all of a sudden if one vehicle fails; adding vehicles has almost no positive influence on delivery time once the number of vehicles exceeds a certain threshold. The timing of job selection has significant impact on delivery time and stability for NJ-policies while it has no relevance for FJ-policies. We also introduced a methodological approach for decision making support to setup a stable delivery system for efficiently resolving the tradeoff between expenditure and service quality. This approach reflects all three time horizons and includes an analytically derived service possibility frontier.

# Acknowledgement

This work was funded in part by ERDF, BABEG, and KWF under grant number KWF-20214/23793/35529. It was carried out in the SOSIE project by Lakeside Labs GmbH.